\documentclass[aps,pra,twocolumn,superscriptaddress,aps]{revtex4}
\usepackage{amsmath,amsfonts,amssymb}

\usepackage{graphicx,graphics}

\begin{document}

\title{Ring dark and anti-dark solitons in nonlocal media}

\author{Theodoros P. Horikis}
\affiliation{Department of Mathematics, University of Ioannina, Ioannina 45110, Greece}

\author{Dimitrios J. Frantzeskakis}
\affiliation{Department of Physics, University of Athens, Panepistimiopolis, Zografos, Athens 15784, Greece}

\begin{abstract}
Ring dark and anti-dark solitons in nonlocal media are found.
These structures have, respectively, the form of annular dips or humps on top of a
stable continuous-wave background, and exist in a weak or strong nonlocality regime,
defined by the sign of a characteristic parameter. It is demonstrated analytically
that these solitons satisfy an effective cylindrical Kadomtsev-Petviashvilli (aka Johnson's)
equation and, as such, can be written explicitly in closed form. Numerical simulations
show that they propagate undistorted
and undergo quasi-elastic collisions, attesting to their stability properties.
\end{abstract}

\maketitle

In nonlinear optics, spatial dark solitons are known to be
intensity dips, with
a phase-jump across the intensity minimum, on top of a continuous-wave (cw)
background beam. These structures may exist in bulk media and waveguides, due to
the balance between diffraction and defocusing nonlinearity, and have been proposed
for potential applications in photonics as adjustable waveguides for weak
signals \cite{kivshar0}.

In the two-dimensional (2D) geometry, spatial dark solitons, in the form of stripes,
are prone to the transverse modulation instability (MI) \cite{kuztur}, which leads to their
bending and their eventual decay into vortices \cite{Ticho}. However, the instability
band of the dark soliton stripes,
may be suppressed if the stripe is bent so as to form a ring
of particular length. This idea led to the introduction of ring dark solitons (RDSs)
\cite{ofyrds}, whose properties have been studied both in theory \cite{djfbam,hector}
and in experiments \cite{RDSexp}, and potential applications of RDS to
parallel guiding of signal beams were proposed \cite{dinev}. RDSs have
also been predicted to occur in other physically relevant contexts, such as
atomic Bose-Einstein condensates \cite{RDSBEC} and polariton superfluids \cite{RDSpol,RDSpolexp}.

While the above results rely on the study of nonlinear Schr\"odinger (NLS) models with a local
nonlinearity, there exist many physical settings where the use of NLS models with a nonlocal
nonlinearity are more appropriate. This occurs, e.g., in media featuring strong thermal
nonlinearity \cite{kroli2} or in nematic liquid crystals \cite{alberucci2}, where the nonlinear
contribution to the refractive index depends on the intensity distribution in the transverse plane.
It has been shown that dark solitons in one-dimensional (1D) settings exist in
media with a defocusing nonlocal nonlinearity
\cite{karta,kong,piccardi,assanto_grey,pu,small_amplitude} while, in the case of stripes,
transverse
MI may be suppressed due to the nonlocality \cite{trillo}.
The smoothing effect of the nonlocal response was shown to occur 
even in the case of shock wave formation \cite{trillo,shock1,shock2,shock3},
or give rise to stable 2D solitons \cite{assanto_book}.
%
%
Here we should note that, generally, pertinent nonlocal models
do not possess soliton solutions in explicit form (other than the weakly nonlinear limit \cite{bang}).
As such, various
techniques have been used to analyze soliton dynamics and interactions, with the most common one
being the variational approximation,
where a particular form of the solution is
chosen \cite{peccianti,alberucci,assanto3,sciberras,alberucci2}. However, to the best of our
knowledge, RDSs in nonlocal media have not been considered so far.

It is the purpose of this article to study 
RDSs and ring anti-dark 
solitons (RASs) in nonlocal media. These structures have, respectively, the form of annular dips or humps
on top of a stable cw background, and exist in a weak or strong nonlocality regime, defined by the
sign of a characteristic parameter. 
Using a multiscale asymptotic expansion technique, we find that RDSs and RASs
obey an effective Johnson's
equation, 
that models ring-shaped waves in shallow water \cite{johnson}. We also
perform direct simulations to show that RDSs and RASs propagate undistorted
and undergo quasi-elastic collisions. 
%

Light propagation in nonlocal media is governed by the following dimensionless model
\cite{karta,alberucci2,peccianti,assanto_book,shock1}:
\begin{subequations}
\begin{gather}
i\frac{{\partial u}}{{\partial z}} + \frac{1}{2}{\nabla ^2}u - 2\eta u = 0,\\
\nu {\nabla ^2}\eta  - 2\eta  =  - 2|u{|^2},
\end{gather}
\label{model}
\end{subequations}
with the transverse Laplacian in cylindrical geometry being:
$\displaystyle{\nabla^2} =
\frac{{{\partial ^2}}}{{{\partial}r^2}} + \frac{1}{r}\frac{\partial }{{\partial r}} +
\frac{1}{{{r^2}}}\frac{{{\partial ^2}}}{{\partial {\theta ^2}}}$.
Here, $u=u(z,r,\theta)$ is the complex electric field envelope, $\eta=\eta(z,r,\theta)$
is the optical refractive index, and the parameter $\nu$ stands for the
strength of nonlocality.
%
%
Notice that two interesting limits are possible: the local limit,
with $\nu$ small, where \eqref{model} reduce to a NLS-type equation with saturable nonlinearity
\cite{reinbert}, and the nonlocal limit, with $\nu$ large. Here, we will treat $\nu$ as
an arbitrary parameter.
%

We start by expressing functions $u$ and $\eta$ as \cite{small_amplitude,horikis}:
\begin{gather*}
u = {u_b}(z)v(r,\theta ,z) = {u_0}{e^{ - 2iu_0^2{\kern 1pt} z}}v(r,\theta ,z), \\
\eta  = {\eta _b}(z)w(r,\theta ,z) = u_0^2 w(r,\theta ,z),
\end{gather*}
where $u_0$ is an arbitrary real constant, while $u_b(z)=u_0 \exp(-2iu_0^2{\kern 1pt} z)$ and
$\eta_b(z) = u_0^2$ form the cw background solution of \eqref{model} so that $v$ and $w$ satisfy:
\begin{subequations}
 \begin{gather}
i\frac{{\partial v}}{{\partial z}} + \frac{1}{2}{\nabla ^2}v - 2\eta_b(w-1) v = 0,\\
\nu{\nabla^2}w  -2 w  =  - 2|v|^2.
\end{gather}
\label{model2}
\end{subequations}
By doing so, we have now fixed constant unit boundary conditions at infinity and the asymptotic
analysis for the determination of $v$ and $w$ may be directly applied. However,
before proceeding further, it is relevant to investigate if the cw background
is subject to MI. We thus perform a standard MI analysis, assuming small perturbations of
$u_b(z)$ and $\eta_b(z)$ behaving like
$\exp[i(k_z z + \boldsymbol{k}_{\perp} \cdot \boldsymbol{r}_{\perp})]$. Then, it is
found that the longitudinal and transverse perturbation wavenumbers $k_z$ and
$\boldsymbol{k}_{\perp}$ obey the dispersion relation:
$k_z^2=2u_0^2 \boldsymbol{k}_{\perp}^2
\left[1+(1/2)\boldsymbol{k}_{\perp}^2\right]^{-1}
+ (1/4)\boldsymbol{k}_{\perp}^4.$
%
%
This equation shows that $k_z$ is always real and, thus, the cw solution
is modulationally stable for the considered model (note that, generally, for certain response functions,
nonlocality could possibly lead to MI even in the defocusing case \cite{kroli}).


Next, we use the Madelung transformation $v=\rho e^{i\phi}$ (where real functions $\rho$ and $\phi$
denote the amplitude and phase of $v$), and obtain from \eqref{model2} the following system:
%
\begin{subequations}
\begin{gather}
  \rho \frac{{\partial \phi }}{{\partial z}} - \frac{1}{2}{\nabla ^2}\rho  + \frac{1}{2}\rho
  {\left(
  {\frac{{\partial \phi
  }}{{\partial r}}} \right)^2} + \frac{1}{{2{r^2}}}{\left( {\frac{{\partial \phi
  }}{{\partial \theta
  }}} \right)^2} +
  2{\eta _b}(w - 1)\rho  = 0, \\
  \frac{{\partial \rho }}{{\partial z}} + \frac{1}{2}\rho {\nabla ^2}\phi  + \frac{{\partial
  \rho
  }}{{\partial
  r}}\frac{{\partial \phi }}{{\partial r}} + \frac{1}{{{r^2}}}\frac{{\partial \rho
  }}{{\partial \theta
  }}\frac{{\partial
  \phi }}{{\partial \theta }} = 0, \\
  \nu{\nabla ^2}w - 2w =  - 2{\rho ^2}.
\end{gather}
\label{model3}
\end{subequations}
Seek, now, small-amplitude solutions on top of the cw background 
in the form of the asymptotic expansions:
\begin{align*}
\rho= \sum_{j=0}^{\infty}{\varepsilon ^{2j}}\rho_{2j}, \quad
\phi=\sum_{j=0}^{\infty}{\varepsilon^{2j+1}}\phi_{2j+1}, \quad
w= \sum_{j=0}^{\infty}{\varepsilon ^{2j}}w_{2j},
\end{align*}
%
%
where the unknown functions depend on the slow variables
$R=\varepsilon(r-Cz)$ (where $C$ is the wave velocity), $\Theta=\theta/ \varepsilon$,
and $Z=\varepsilon^3 z$.
Substituting these expansions to \eqref{model3} we obtain a hierarchy of coupled systems.
To leading order in $\varepsilon$, i.e., for $O(\varepsilon^{-4})$
and $O(\varepsilon^{-3})$, a system of linear equations is obtained:
\begin{gather}
\left. {\begin{array}{c}
\displaystyle{2u_0^2{w_2} - C\frac{{\partial {\phi _1}}}{{\partial R}} = 0,} \\[6pt]
\displaystyle{ - 2C\frac{{\partial {\rho _2}}}{{\partial R}}
+ \frac{{{\partial ^2}{\phi _1}}}{{\partial {R^2}}} = 0,}
\\[3pt]
\displaystyle{{w_2} = 2{\rho _2},}
\end{array}} \right\} \Leftrightarrow \left\{ {\begin{array}{c}
\displaystyle{\frac{{\partial {\phi _1}}}{{\partial R}} = \frac{{4u_0^2}}{{C}}{\rho _2}}, \\[6pt]
\displaystyle{C^2} = 2u_0^2,\\[3pt]
\displaystyle{{w_2} = 2{\rho _2}}.
\end{array}} \right.
\label{velocity}
\end{gather}
Notice that the velocity $C$, determined by \eqref{velocity}, may have two
signs, corresponding to outward or inward propagating ring solitons (see below).
Next, at $O(\varepsilon^{1})$ and $O(\varepsilon^{-2})$ we get:
\begin{gather}
  2{C^2}{w_4} + 4{C^2}\rho _2^2 + 2\frac{{\partial {\phi _1}}}{{\partial Z}} -
  2C\frac{{\partial {\phi
  _3}}}{{\partial R}}
  - \frac{{{\partial ^2}{\rho _2}}}{{\partial {R^2}}} = 0,
  \label{w4}\\
  {w_4} = \rho _2^2 + 2{\rho _4} + \nu\frac{{{\partial ^2}{\rho _2}}}{{\partial
  {R^2}}},
  \label{rho4}
\end{gather}
and at $O(\varepsilon^{-1})$:
\begin{gather*}
 2{C^2}{Z^2}\frac{{\partial {\rho _2}}}{{\partial Z}} + \frac{{{\partial ^2}{\phi
 _1}}}{{\partial
 {\Theta ^2}}} -
 4{C^2}RZ\frac{{\partial {\rho _2}}}{{\partial R}} - 2{C^3}{Z^2}\frac{{\partial {\rho
 _4}}}{{\partial
 R}}+
 2CRZ\frac{{{\partial ^2}{\phi _1}}}{{\partial {R^2}}}\nonumber \\
   + {C^2}Z{\rho _2}\left( {2 + {\text{4}}CZ\frac{{\partial {\rho _2}}}{{\partial R}} +
   Z\frac{{{\partial ^2}{\phi
   _1}}}{{\partial {R^2}}}
   + {C^2}{Z^2}\frac{{{\partial ^2}{\phi _3}}}{{\partial {R^2}}}} \right) = 0.
\end{gather*}

\begin{figure}[t]
\centering
\includegraphics[height=4cm]{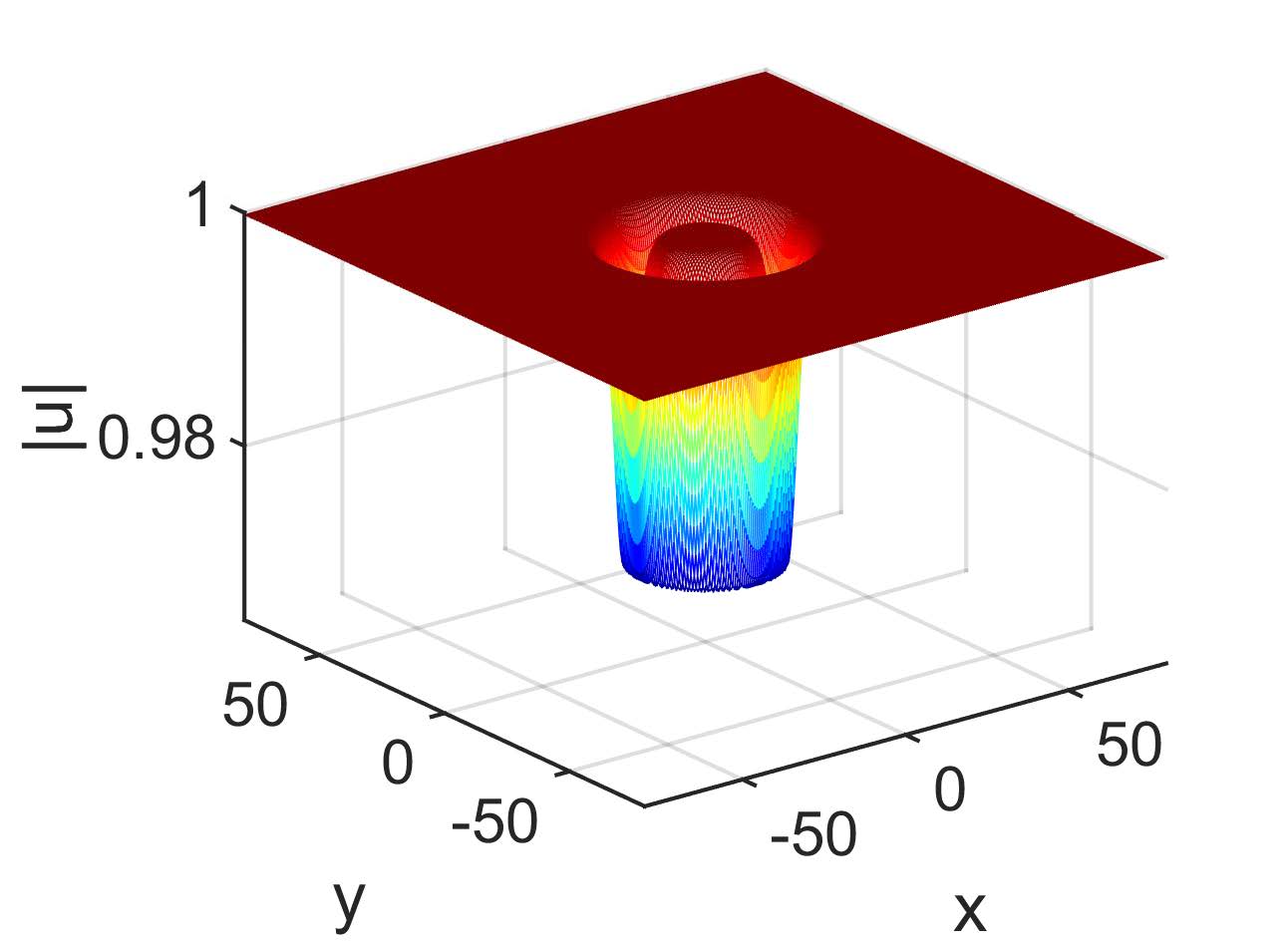}\\
\includegraphics[height=4cm]{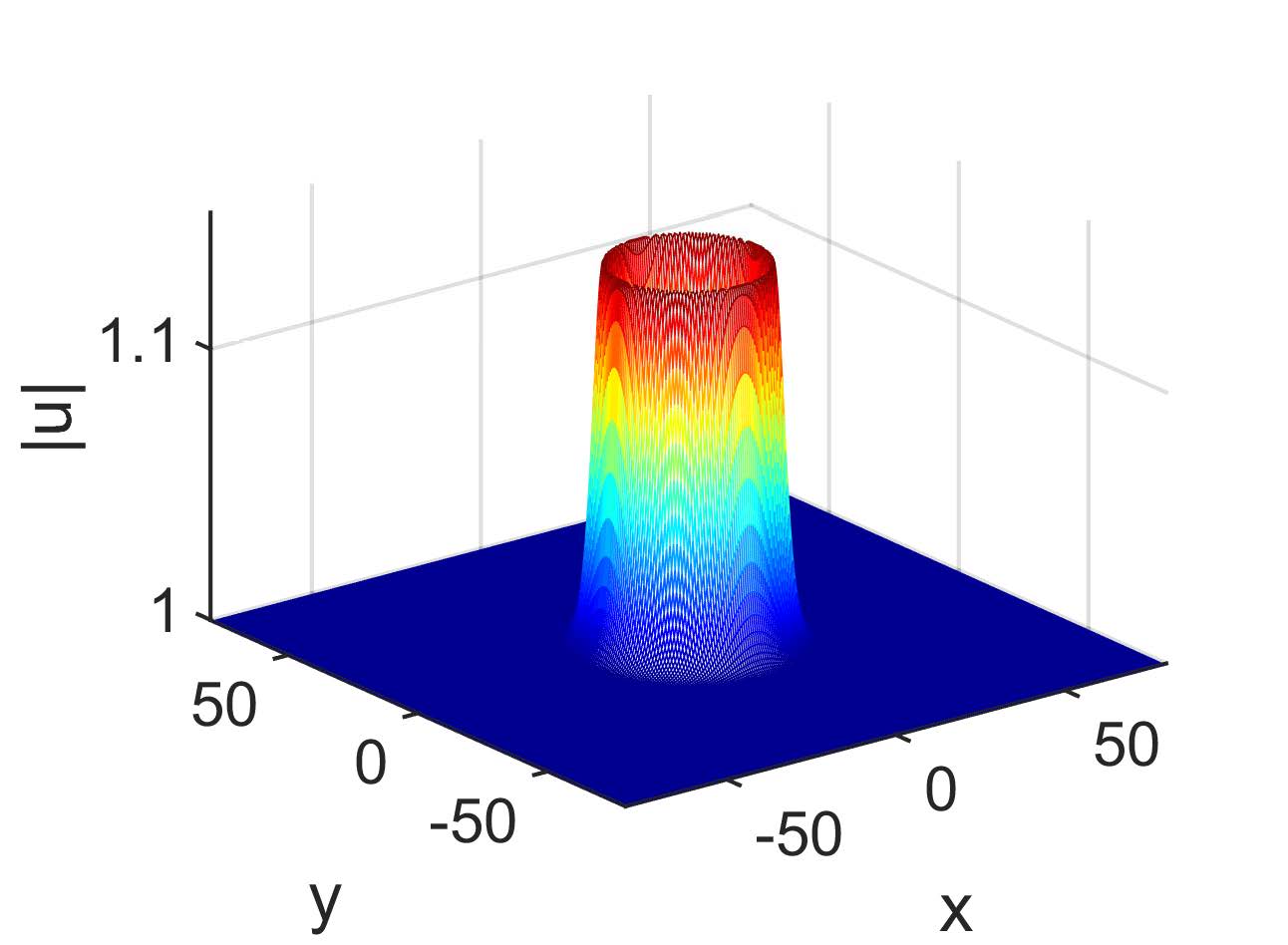}
\caption{(Color Online) Typical RDS (top) and RAS (bottom),
for $u_0=\alpha=1$ and $\nu=1$ ($\nu=1/3$) for the RAS (RDS).}
\label{soliton3D}
\end{figure}

Next, solve \eqref{w4}-\eqref{rho4} for $\rho_4$ and substitute above to
obtain the following nonlinear evolution equation for $\rho_2$:
\begin{gather}
\frac{\partial}{\partial R} \left(
\frac{\partial\rho _2}{\partial Z} +
\frac{3C}{2}\rho_2 \frac{\partial\rho _2}{\partial R} + \frac{SC}{8}
\frac{\partial ^3\rho _2}{\partial R^3} +
\frac{1}{2Z}\rho _2
\right)
+ \frac{1}{2CZ^2}\frac{\partial ^2\rho _2}{\partial\Theta^2} = 0,
\label{ckp}
\end{gather}
where parameter $S$ is given by:
\[
S=\frac{2C^2\nu-1}{C^2} = \frac{4 u_0^2 \nu-1}{2u_0^2}.
\]
%
Equation (\ref{ckp}) is a cylindrical Kadomtsev-Petviashvili (cKP) equation, also known as
Johnson's equation, first introduced in the context of shallow water waves \cite{johnson}.
There exist transformations \cite{klein} linking this model with the more commonly known
KP equation in the Cartesian geometry \cite{ist}, which allows for construction of solutions
of cKP from solutions of KP. Although ---obviously--- there exist other choices, here
we focus on solutions with radial symmetry, which do
not depend on $\Theta$. In this case, the system reduces to the cylindrical Korteweg-de Vries
(cKdV) equation, which possesses cylindrical, sech$^2$-shaped soliton solutions,
on top of a rational background \cite{hirota}:
%
%
\begin{gather}
\rho_2(R,Z) = \frac{R}{3CZ} + \frac{S\alpha ^2}{Z}\mathrm{sech}^2\left( \frac{S C\alpha ^3}{\sqrt{Z}}
+ \frac{\alpha R}{\sqrt Z }  + R_0 \right),
\label{soliton}
\end{gather}
where, $\alpha$ is an arbitrary real parameter [of order $O(1)$].
Note that the characteristics of the solitons' core, i.e., amplitude, power, velocity,
and inverse width, scale as:
$\alpha^2$, $\alpha^4$, $\alpha^2$, and $\alpha$, respectively, similarly to
the case of the usual KdV solitons \cite{ist}.

Clearly, the sign of parameter $S$
%
determines the nature of the soliton: if $S<0$, the solitons are depressions off of the cw
background and are, hence, dark solitons; if $S>0$, the solitons are humps on top of the cw
background and are, thus, anti-dark solitons (note that if $S \rightarrow 0$,
modification of the asymptotic analysis and inclusion of higher-order terms is needed as,
e.g., in the shallow water wave problem \cite{burde}).
Examples of these RDS and RAS solutions,
as introduced above,
are shown in Fig.~\ref{soliton3D}. Notice that, having determined the form of the soliton
[\eqref{soliton}], the refractive index can readily be found in terms of $\rho_2$: in fact,
up to $O(\epsilon^2)$, it is given by $n=n_b w \approx u_0^2(1+2\epsilon^2 \rho_2)$, thus
having the form of an annular well (barrier) for the RDS (RAS).

%

Here, recalling that $\nu$ defines the degree of
nonlocality, it is important to observe
that $S<0\Rightarrow \nu < (1/2u_0)^2$, while $S>0\Rightarrow \nu >
(1/2u_0)^2$. These inequalities indicate that RDS (RAS) are supported in a regime of weak (strong)
nonlocality, as defined by the sign of $S$. Indeed, in the local limit with $\nu=0$ the NLS does
not exhibit these RASs.


\begin{figure}[t]
\centering
\includegraphics[height=3cm]{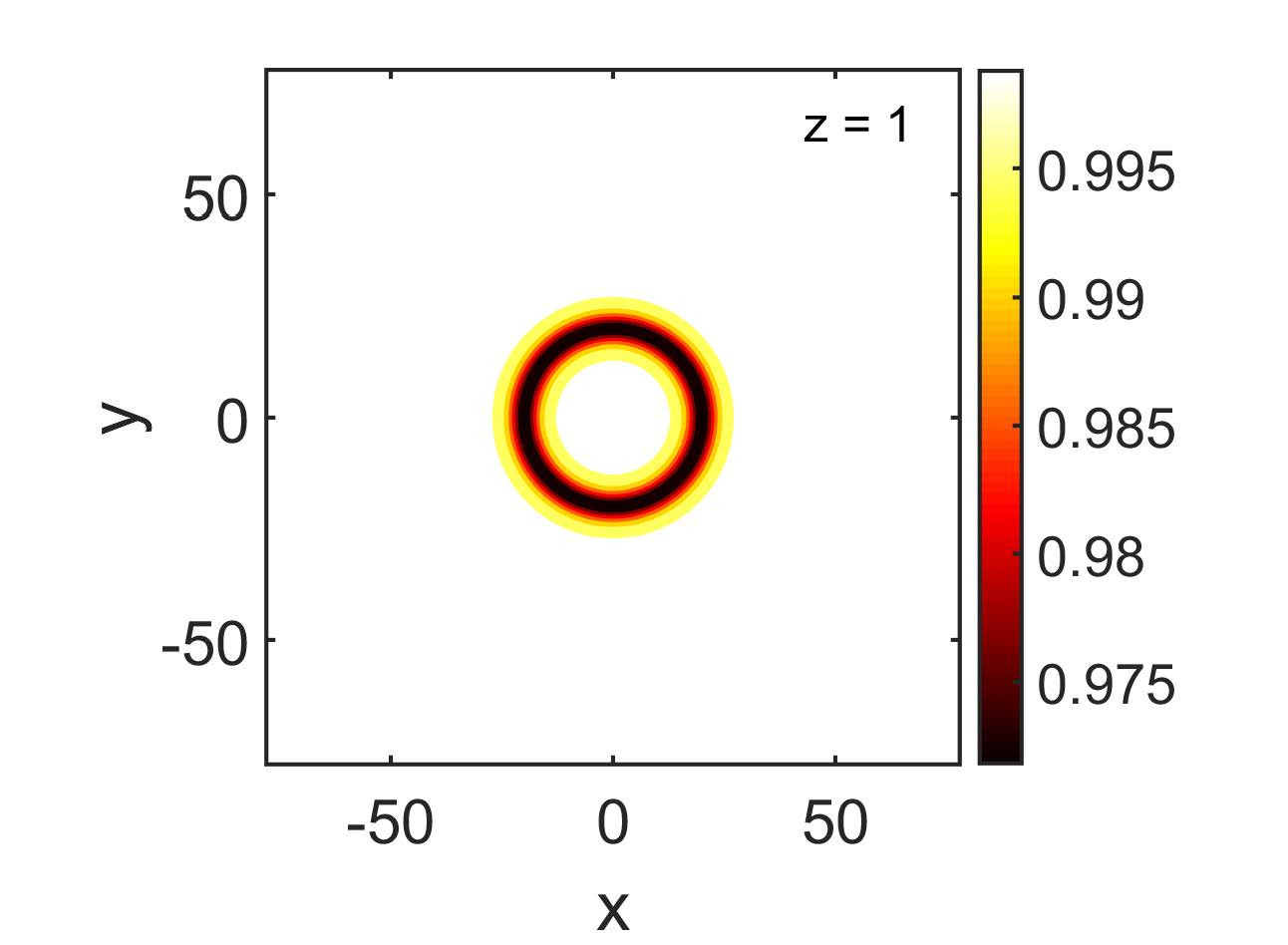}
\includegraphics[height=3cm]{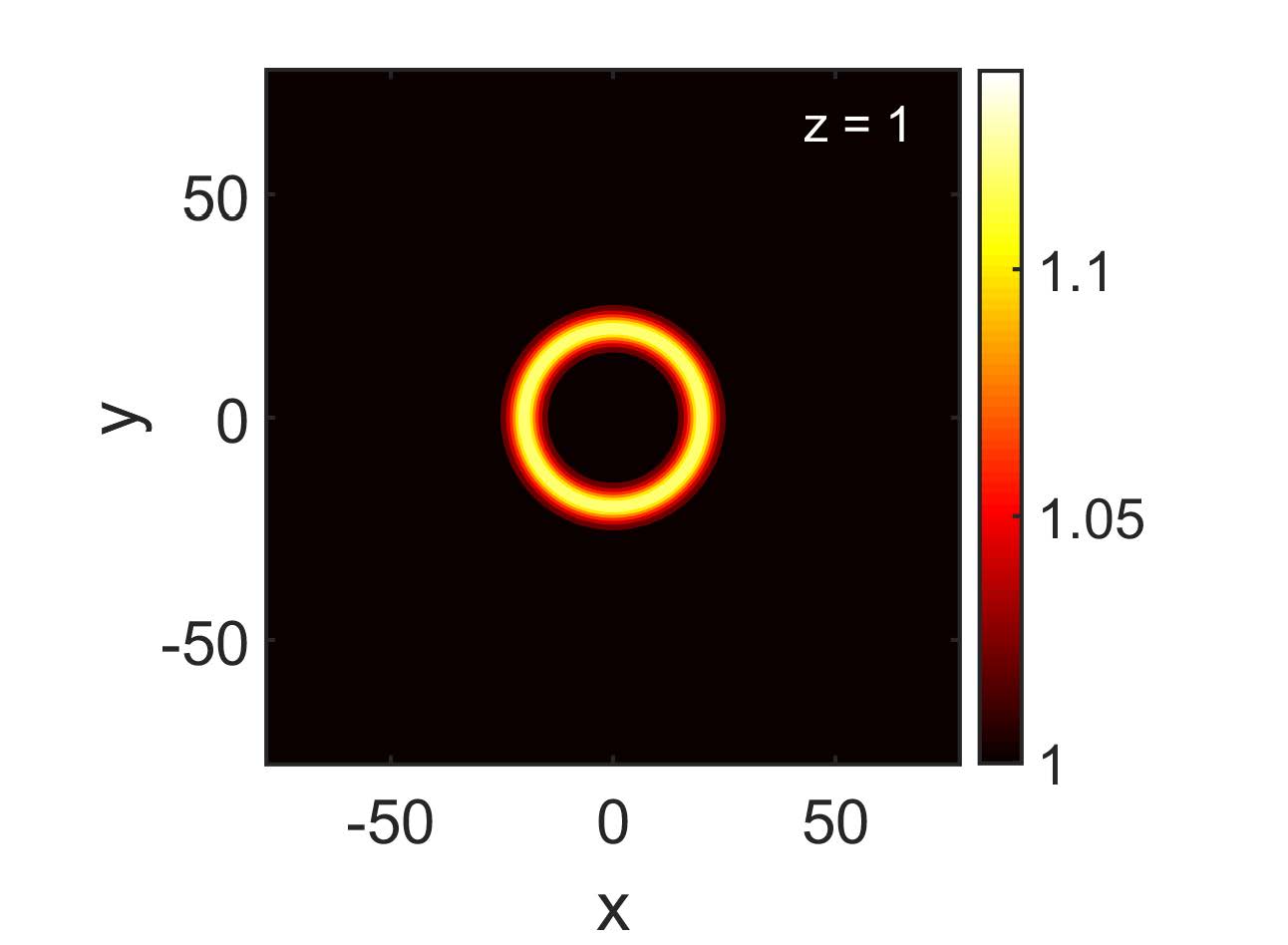}\\
\includegraphics[height=3cm]{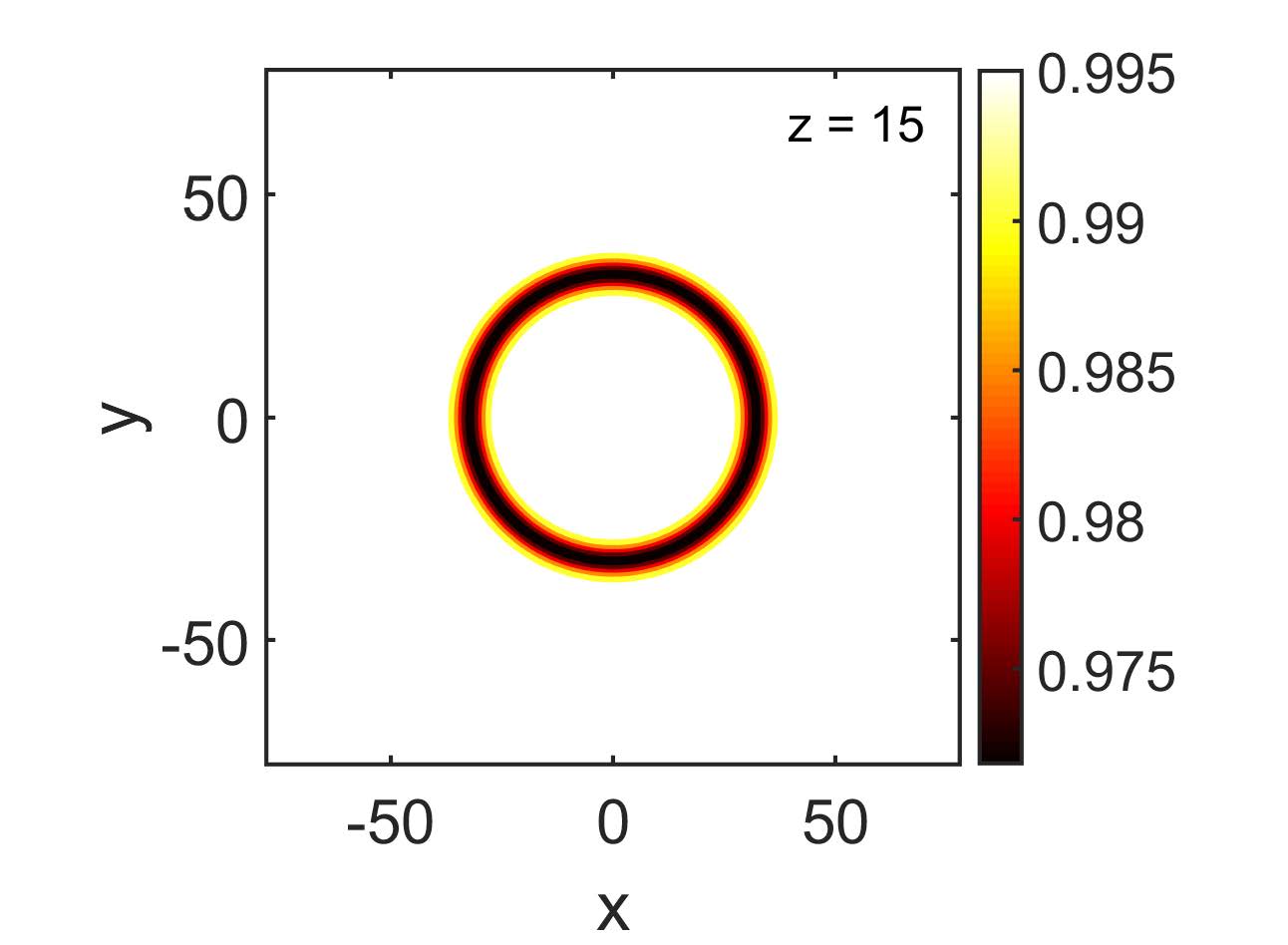}
\includegraphics[height=3cm]{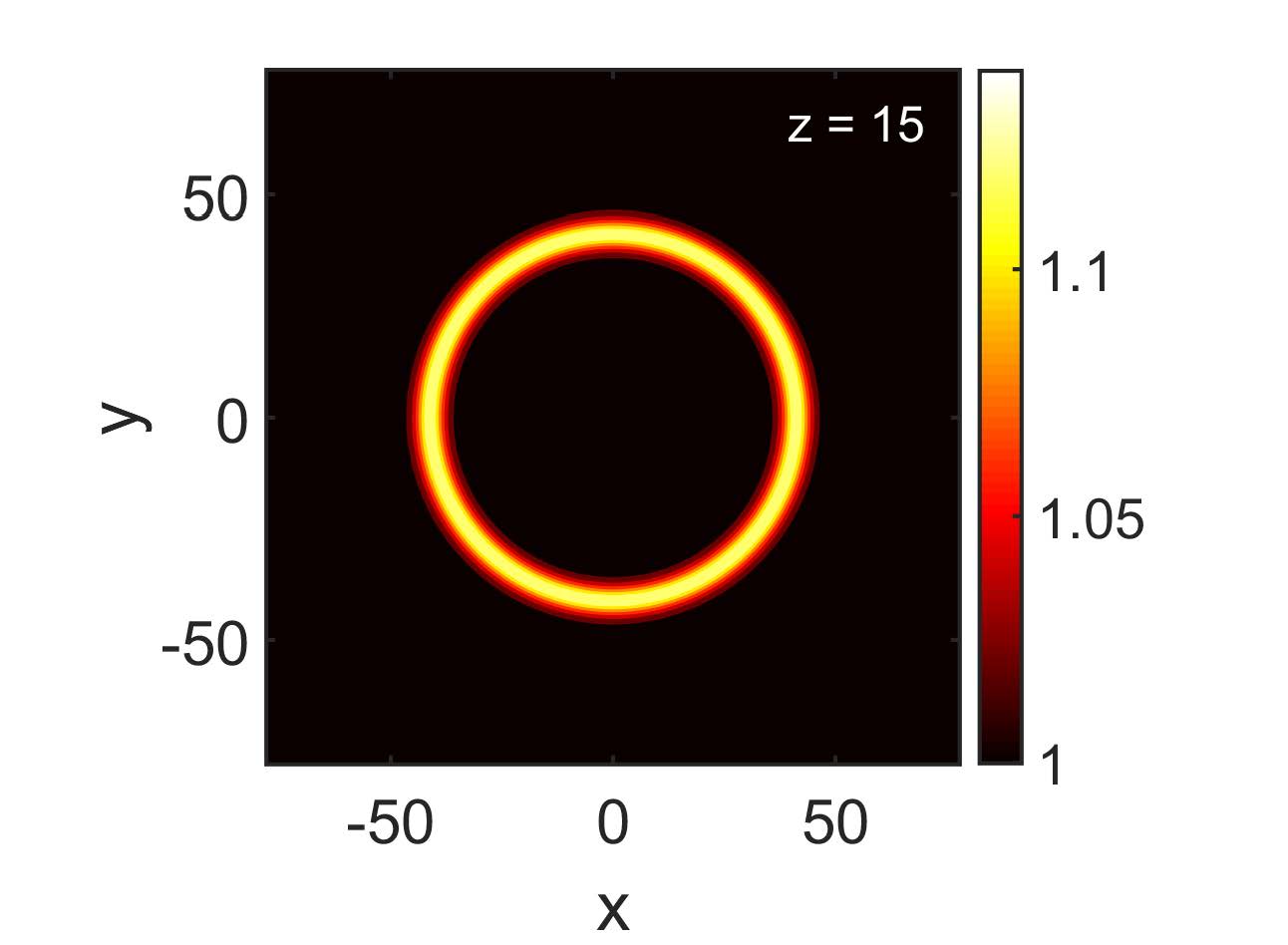}
\caption{(Color Online) Evolution of a RDS (left) and a RAS (right).
Parameter values are as in Fig.~\ref{soliton3D}.}
\label{contours}
\end{figure}

To numerically investigate the propagation properties of RDS and RAS, we evolve an initial ($z=1$)
profile for both cases. Note that the rational background is not shown in Cartesian coordinates
---see Ref.~\cite{santini} for a discussion on the asymptotics for $z\rightarrow 0$. In
Fig.~\ref{contours}, we evolve this initial condition under \eqref{model} for $u_0=\alpha=1$ and
$\nu=1$ ($\nu=1/3$) for the RAS (RDS); note that in the simulations we use a high accuracy
spectral integrator in Cartesian coordinates. 
We find that the role of nonlinearity is crucial for the soliton formation:
indeed, in the linear regime, the electric field envelope features a
diffraction-induced broadening, while when nonlinearity is present a strong localization
is observed (results not shown here), and solitons are formed.
Other interesting features, directly connected with the soliton form,
\eqref{soliton},
are reported below.

First, the two solitons propagate undistorted, i.e., the initial rings expand outwards, keeping
their shapes during the evolution -- at least for relatively short propagation distances.
This fact, however, does not ensure stability of solitons, especially against azimuthal
perturbations. Nevertheless,
information regarding the RDS and RAS stability
can be inferred from the cKP: in fact, \eqref{ckp} includes both models, so-called \cite{klein}
cKP-I (for $S<0$) and cKP-II (for $S>0$). Then, similarly to the case of the KP equation,
where lower-dimensional line (KdV-type) solitons of KP-I (KP-II) are unstable
(stable) against transverse perturbations \cite{ist}, we can infer the following: 
ring (cKdV-type) solitons of cKP-I (cKP-II),
i.e., the RDS (for $S<0$) and RAS (for $S>0$) respectively,
are unstable (stable) against azimuthal
perturbations. Nevertheless, in our simulations we have not observed the instability of RDS,
for propagation distances up to $z\approx 70$.

Second, we find that the soliton velocities are
$C=\sqrt{2/3}$ for the RDS and $C=\sqrt{2}$ for the RAS, with a
deviation less than $2\%$ from the analytical prediction.
It is also observed in Fig.~\ref{contours} that, indeed, the RAS's radius
is larger than that of RDS at the same propagation distance.
Note that the amplitudes of RDS and RAS depend on $S$, but do not depend
on $C$ (the sign of $C$ determines if the soliton will contract inwards or expand outwards). Thus,
according to these results,
the RDS and the RAS cannot coexist.
However, we note that in the presence of a competing quintic nonlinearity, it would be
in principle possible to find parameter regimes where RDS and RAS do coexist, as was the
case in Refs.~\cite{saturable1,saturable2} (see also Refs.~\cite{kivshar3,kivaf,difdim} for
the same effect in a setting incorporating third-order dispersion).

\begin{figure}[ht]
\centering
\includegraphics[height=3cm]{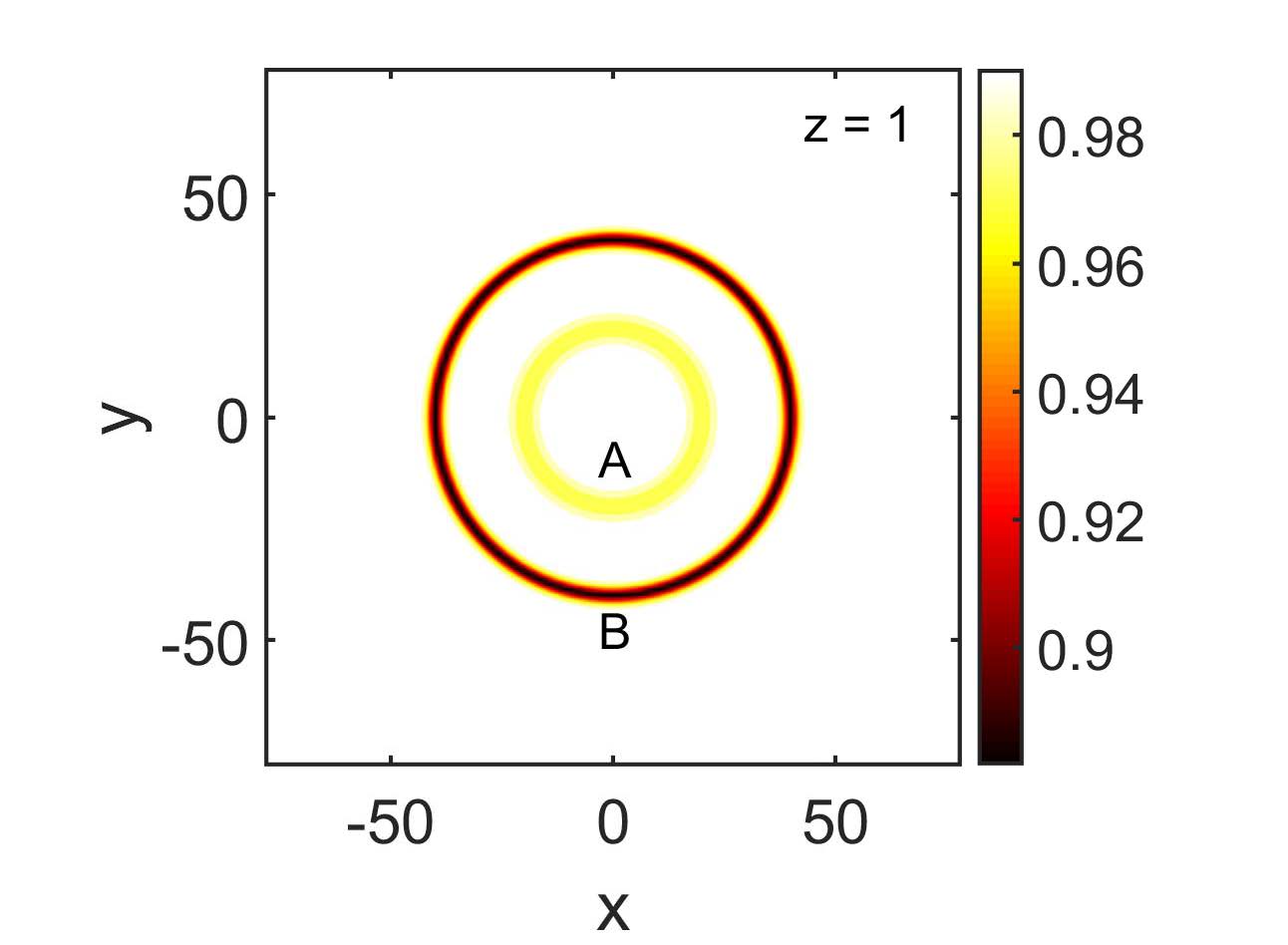}
\includegraphics[height=3cm]{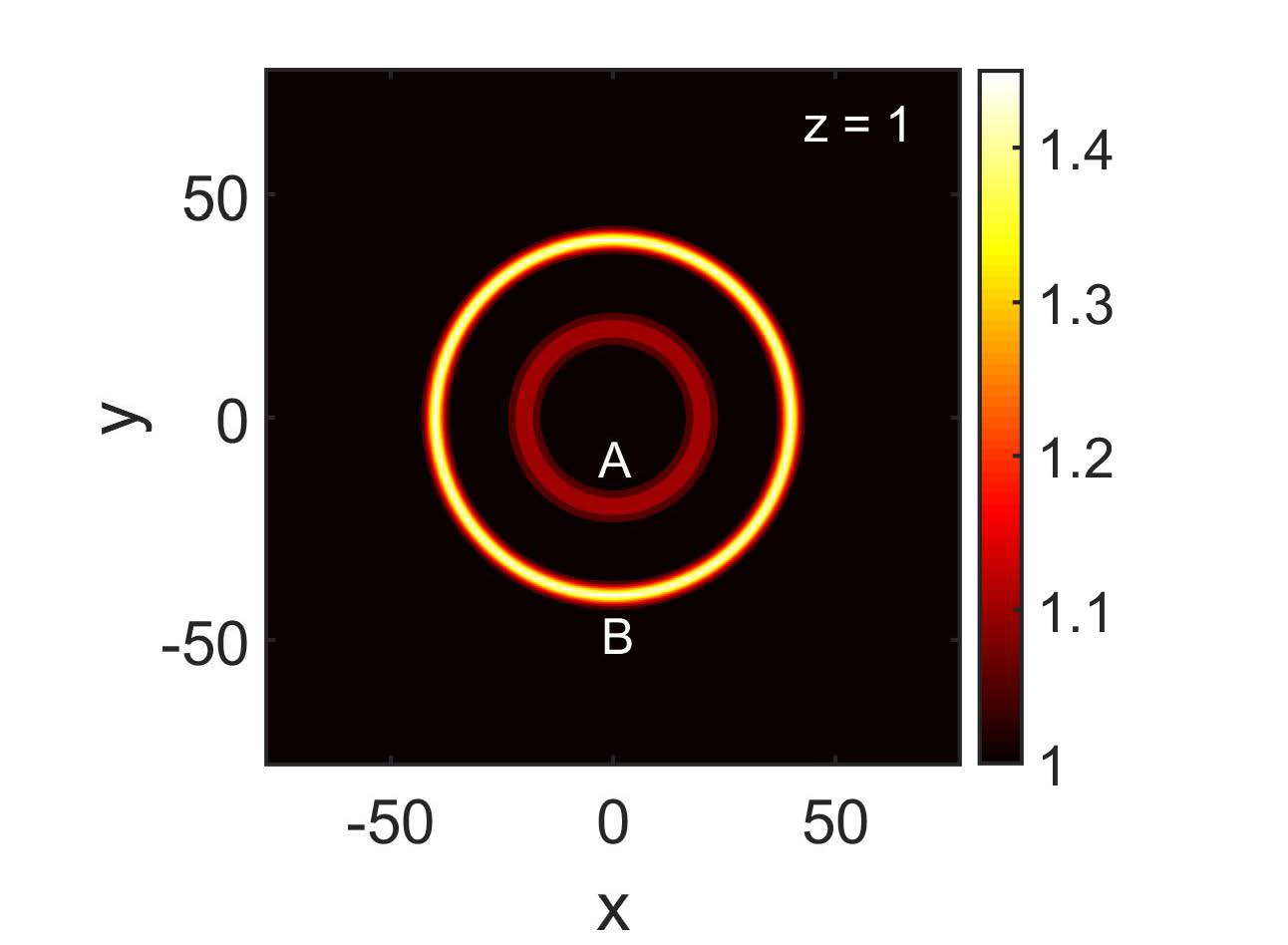}\\
\includegraphics[height=3cm]{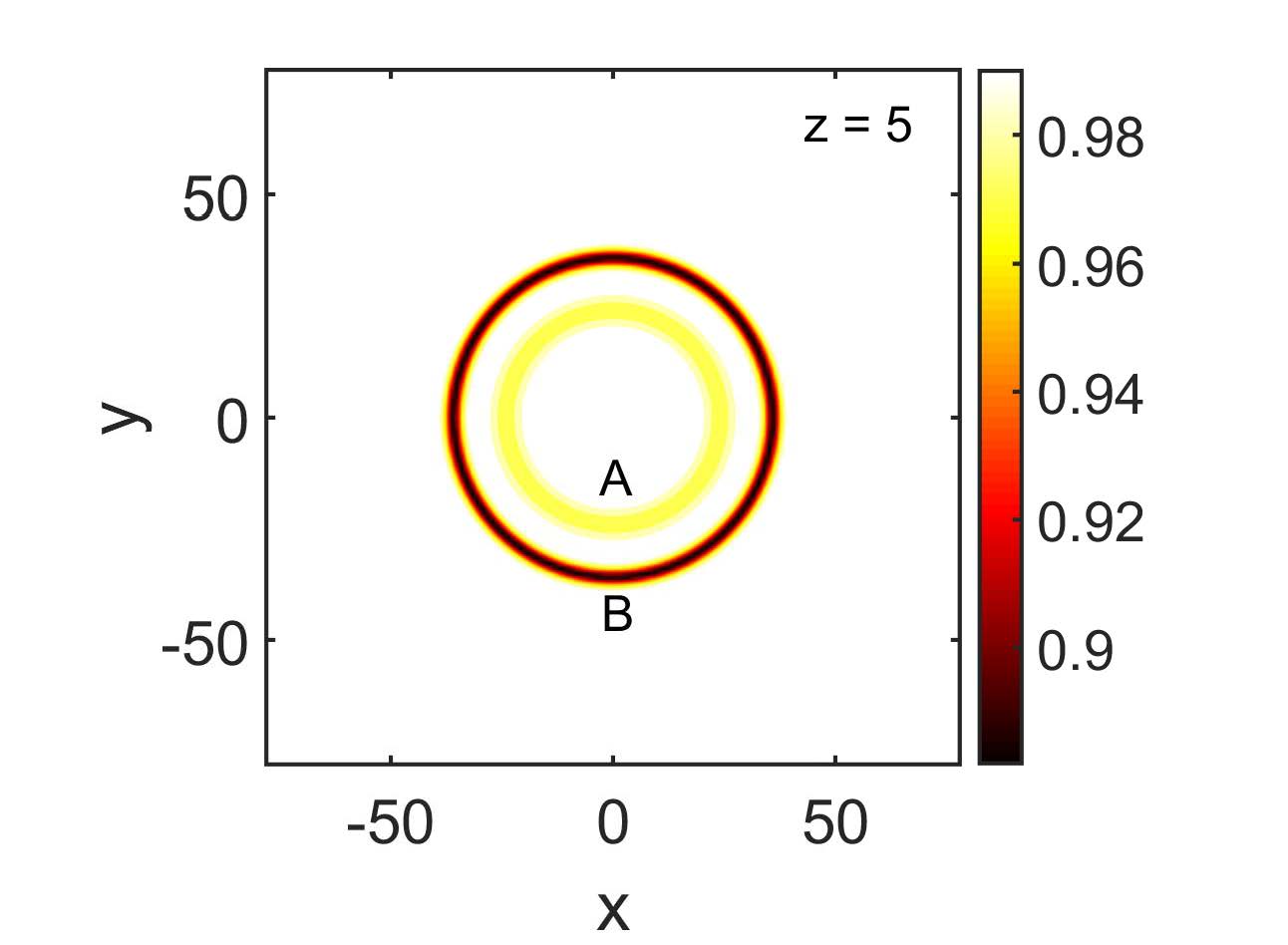}
\includegraphics[height=3cm]{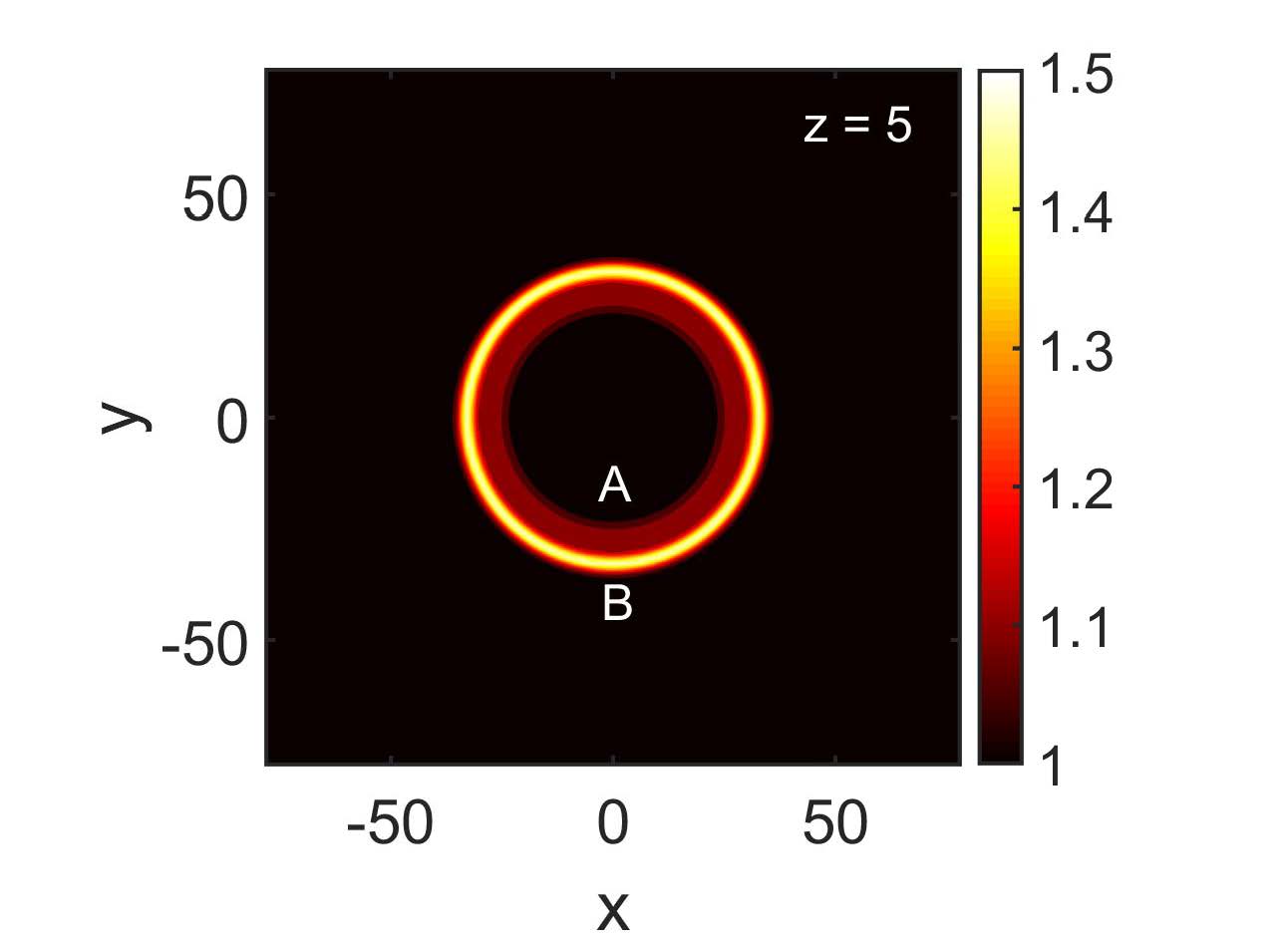}\\
\includegraphics[height=3cm]{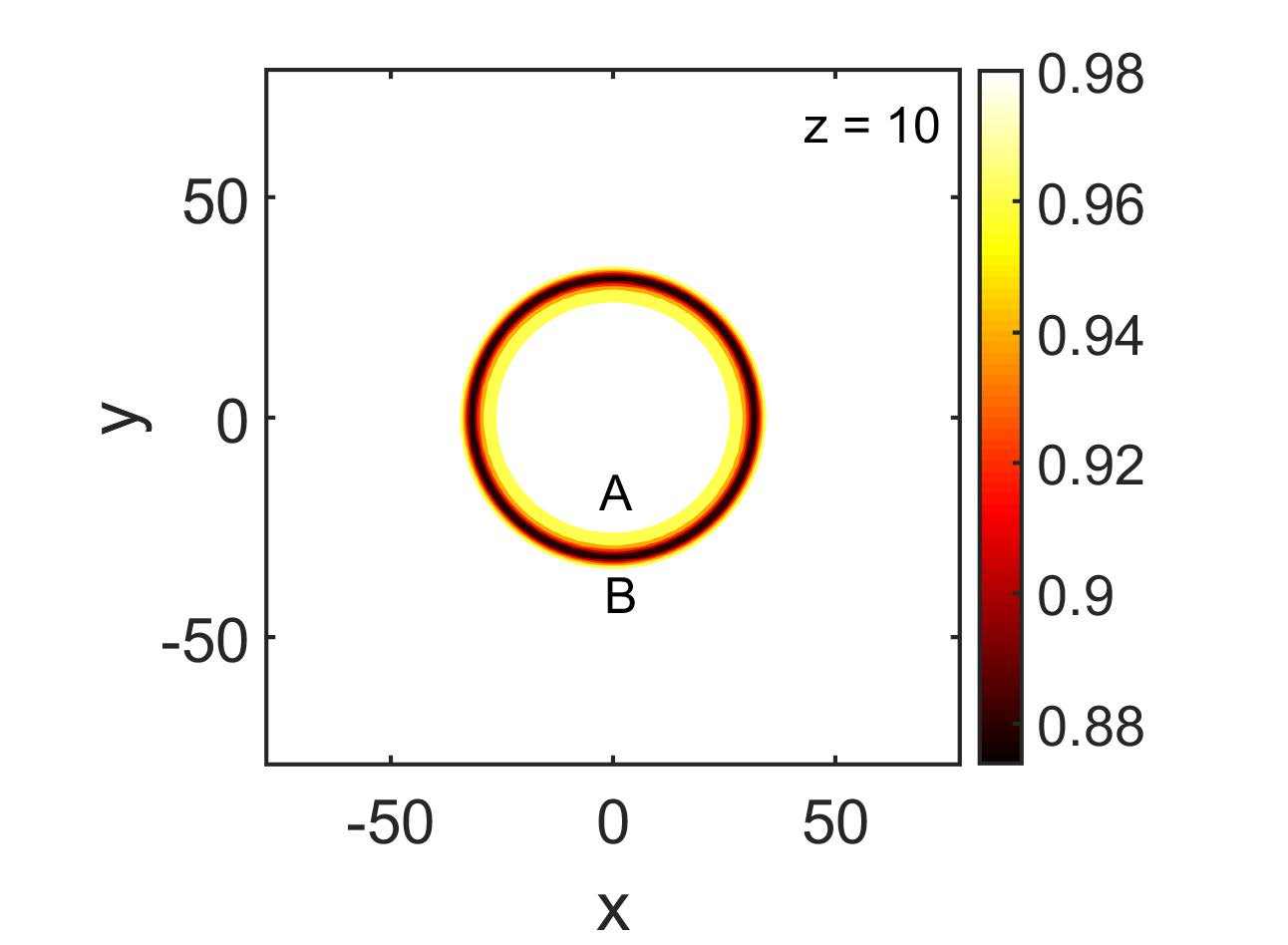}
\includegraphics[height=3cm]{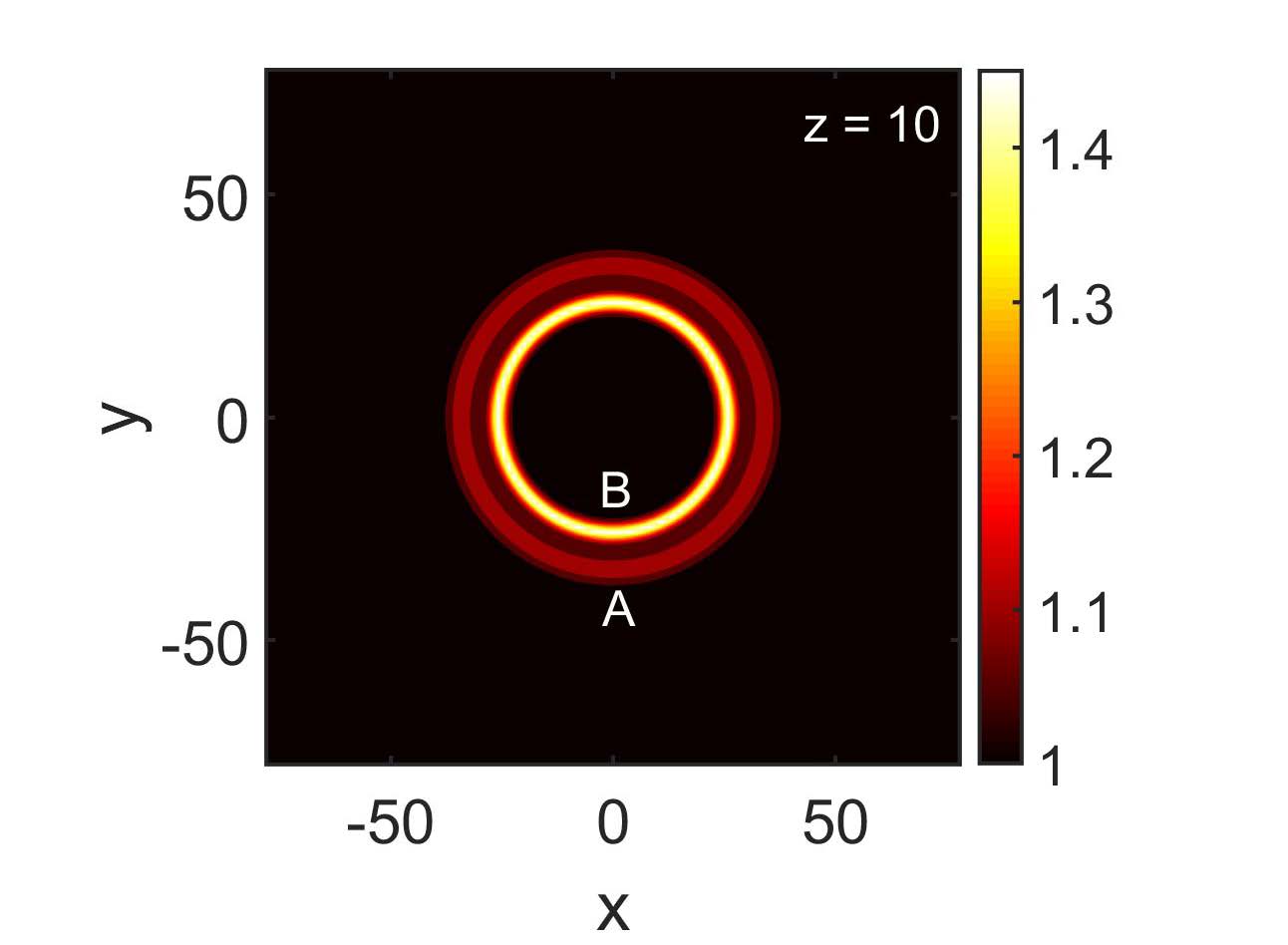}\\
\includegraphics[height=3cm]{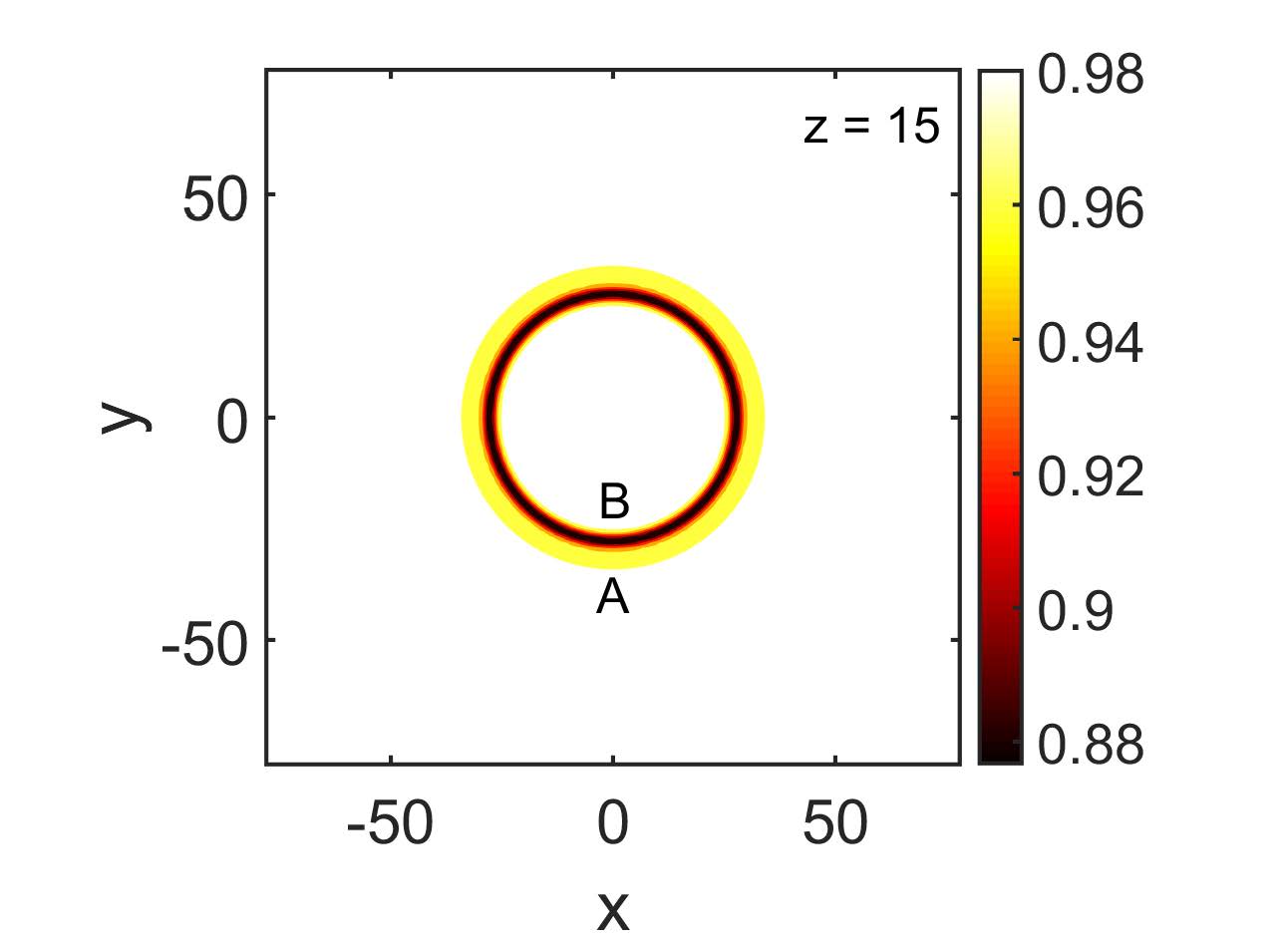}
\includegraphics[height=3cm]{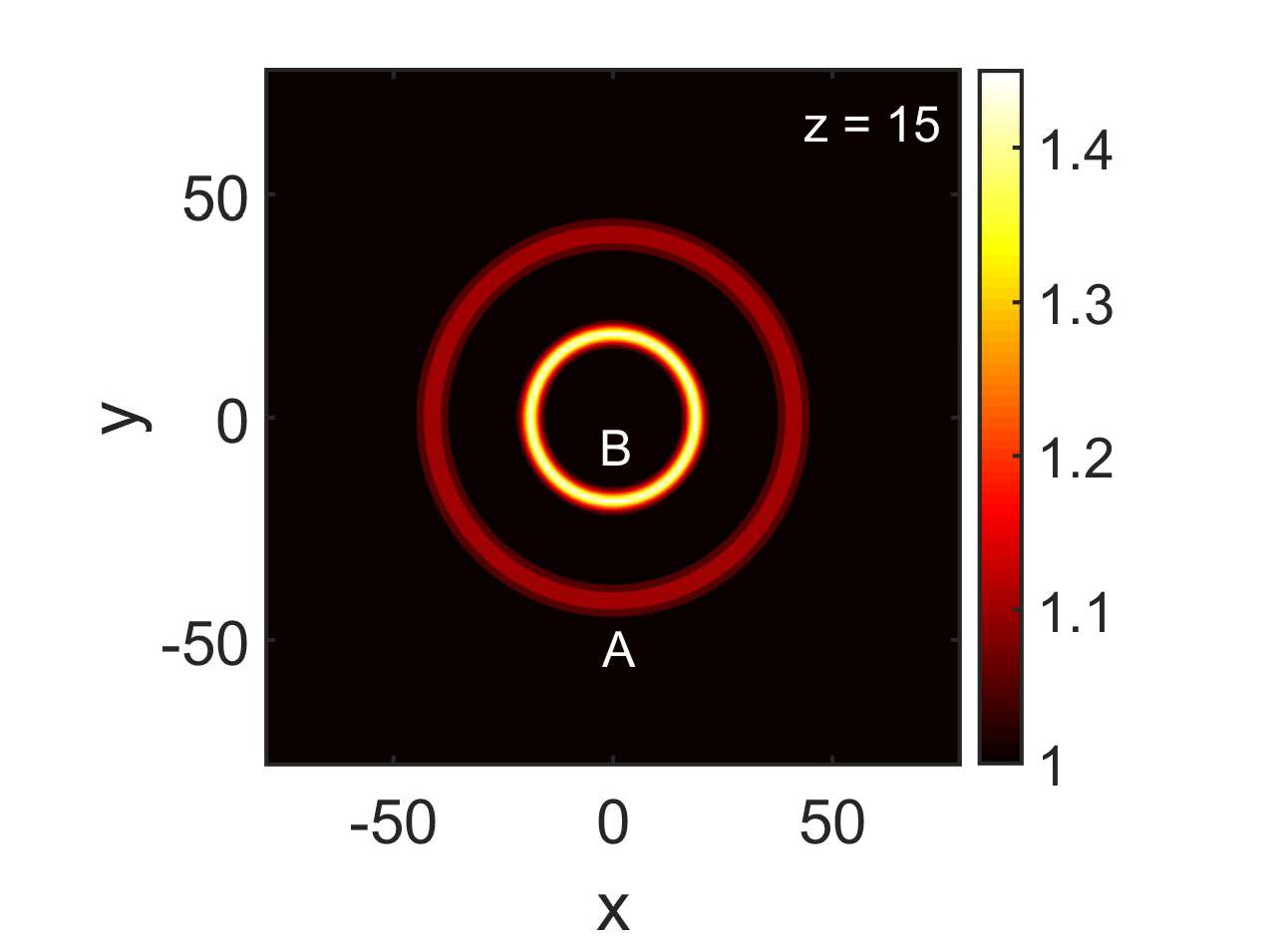}\\
\includegraphics[height=3cm]{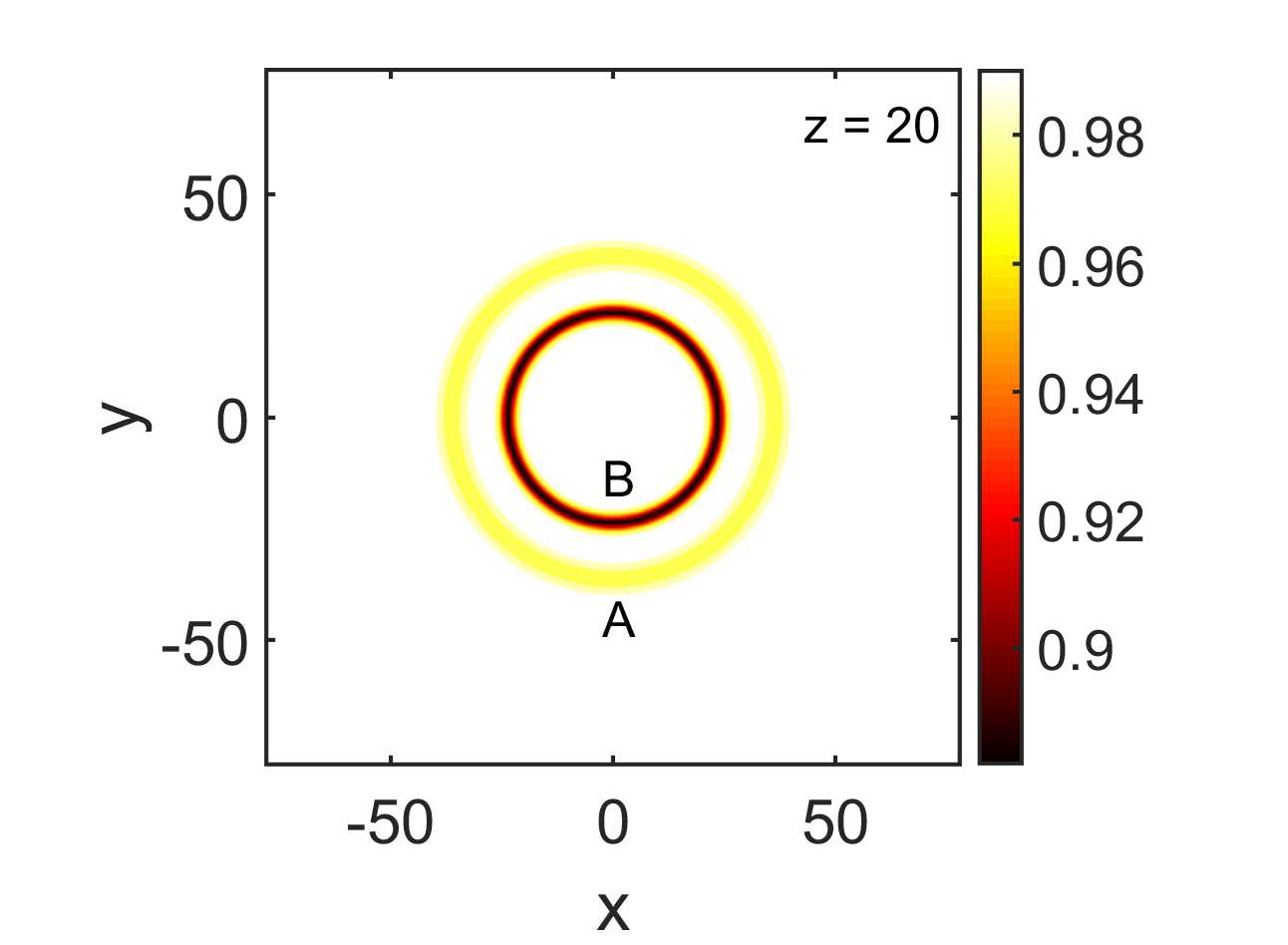}
\includegraphics[height=3cm]{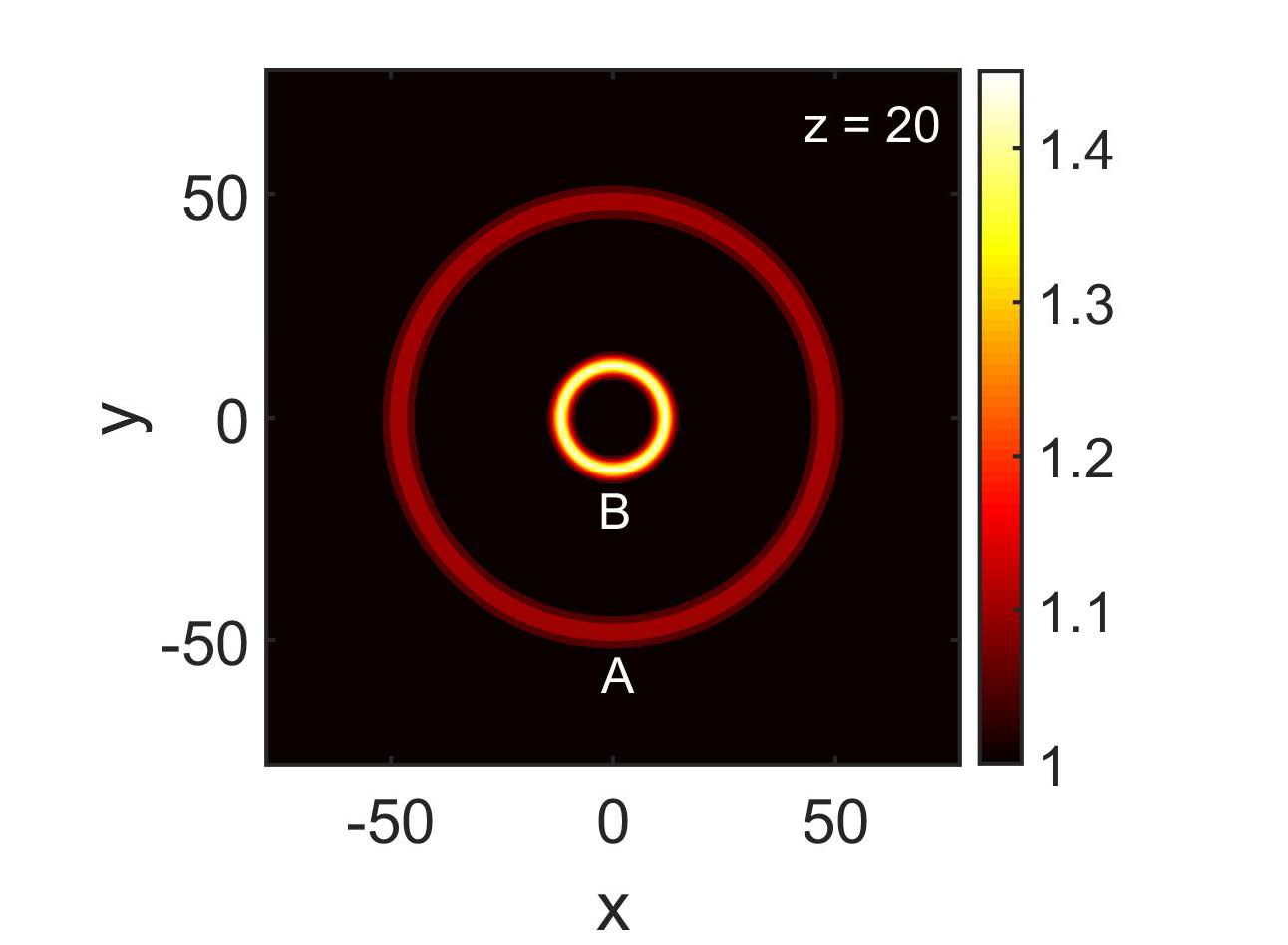}
\caption{(Color Online) Interaction of RDSs (left) and RNAs (right) solitons.}
\label{contours2}
\end{figure}

Third, although RDS and the RAS cannot coexist and, thus, 
cannot interact with each other,
it is possible to study interactions of two solitons of the same type, namely RDS-RDS or RAS-RAS:
this would be an important test on their robustness and solitonic character ---at
least up to relatively short propagation distances, as explained above.
In Fig.~\ref{contours2}, we show the interaction of two solitons of unequal amplitudes, namely
$\alpha=1$ ($\alpha=2$) for the inner (outer) soliton; other parameters are as above. The
velocities are also chosen as before, but with a different sign, so that the solitons will undergo a
head-on collision. As seen, the collision is quasielastic: after passing through each other, the solitons
restore their shapes. This behavior is in agreement with the perturbation theory of
Ref.~\cite{hector}, which predicts that the head-on collision is elastic up to the second-order.

It should be mentioned that the numerical results obtained above refer to the collision between
concentric RDS or RAS. However, there exists the possibility of the collision between slightly
mismatched rings. In this case, it is expected that the collision will produce small oscillations
of the rings, that will be oscillating between two elliptic configurations with small positive and
negative eccentricities.

To conclude, we have found and analyzed ring dark solitons (RDSs) and ring
anti-dark solitons (RASs) in nonlocal media.
These structures were found as special, radially symmetric, solutions of a cylindrical KdV model,
which is a lower-dimensional reduction of an underlying
cylindrical KP (alias Johnson's) equation. RDSs and RASs are supported, respectively,
in a weak or strong nonlocal regime, as defined by the sign of a characteristic parameter.
The same parameter controls the stability of these structures:
in particular, RDSs (RASs) are predicted to be unstable (stable) against azimuthal perturbations.
These facts highlight the role of nonlocality, which, not only support RASs that do not exist
in the local limit, but also renders them stable in the higher-dimensional setting.
For relatively short propagation distances, both structures were found to propagate
undistorted
and remain unaffected even under head-on collisions; this attests to their solitonic character.
Importantly, even for longer
propagation distances, instabilities were not observed in our simulations. This suggests
that RDSs and RASs have a good chance to be observed in experiments.

\bibliography{biblio_nematicons}

\end{document}